\documentclass[aps,pra,floatfix,preprint,showpacs]{revtex4-1}

\usepackage{graphicx}
\usepackage{bm}
\usepackage{amssymb}
\usepackage{subfigure}

\begin{document}


\title{Effect of Strong Electron Correlation on the Efficiency of Photosynthetic Light Harvesting}

\author{David A. Mazziotti}

\email{damazz@uchicago.edu}
\affiliation{Department of Chemistry and The James Franck Institute, The University of Chicago, Chicago, IL 60637}%

\date{Citation: D. A. Mazziotti, J. Chem. Phys. {\bf 137}, 074117 (2012)}


\begin{abstract}

Research into the efficiency of photosynthetic light harvesting has
focused on two factors: (1) entanglement of chromophores, and (2)
environmental noise.  While chromophores are conjugated
$\pi$-bonding molecules with strongly correlated electrons, previous
models have treated this correlation implicitly without a
mathematical variable to gauge correlation-enhanced efficiency. Here
we generalize the single-electron/exciton models to a
multi-electron/exciton model that explicitly shows the effects of
enhanced electron correlation within chromophores on the efficiency
of energy transfer. The model provides more detailed insight into
the interplay of electron correlation within chromophores and
electron entanglement between chromophores. Exploiting this
interplay is assisting in the design of new energy-efficient
materials, which are just beginning to emerge.

\end{abstract}

\pacs{31.10.+z}

\maketitle

\section{Introduction}

Nature harvests solar energy with a remarkably high {\em quantum
efficiency}, the percentage of charge carriers created by photons.
Recent spectroscopic experiments~\cite{ECR07,CWW10,PHF10} and
theoretical models~\cite{SM11,CS09,CCD10,HC10,MN11,
PAM09,ZKR11,BWV10,BE11,SIF10}, provide evidence that efficient light
harvesting in nature occurs by a quantum mechanism involving
sustained electronic coherence~\cite{BCR10} and
entanglement~\cite{K07b,HHH09} between chromophores.  While the
chromophores are chlorophyll molecules containing large networks of
conjugated carbon bonds that surround a charged magnesium ion, they
have largely been represented in theoretical
studies~\cite{SM11,CS09,CCD10,HC10,MN11,PAM09,ZKR11,BWV10,BE11,SIF10}
by one-electron models that neglect the effects of electron
correlation and entanglement within chromophores.  Two advanced
methods in electronic structure, density-matrix renormalization
group~\cite{HDA07} and two-electron reduced-density-matrix
theory~\cite{M11,E07}, have recently shown that networks of
conjugated bonds as in acene chains~\cite{HDA07,GM08,*PGG11}, acene
sheets~\cite{GM08,*PGG11}, and chlorophyll are associated with
polyradical character that cannot be adequately described without a
strongly correlated many-electron quantum model.

\begin{figure}[ht!]

\label{fig:lip}

\begin{center}
\includegraphics[width=2.50in, height=1.5in]{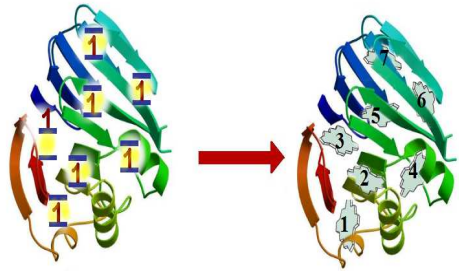}
\end{center}

\caption{{\bf Single electrons or correlated chromophores.} Each of
the seven chromophores in the FMO complex is generally treated as a
single electron in a two-state model (left), and yet the
chromophores are constructed from chlorophyll molecules with many
strongly correlated electrons (right).  Here we treat each of the
chromophores by a correlated $N$-electron model by Lipkin, Meshkov,
and Glick.  Illustration by K. Naftchi-Ardebili, The University of
Chicago, 2011.  Used with permission.}

\end{figure}

In this paper we examine the efficiency of light harvesting where we
represent each chromophore by a correlated $N$-electron model to
treat strong electron correlation.  Figure~1 illustrates the
replacement of one-electron models for each of the 7 chromophores in
the Fenna-Matthews-Olson (FMO) complex of green-sulfur bacteria by
$N$-electron models of increasing complexity.  Here we employ the
$N$-electron Lipkin-Meshkov-Glick (LMG) model~\cite{LMG65,MOP06}
used extensively in electronic structure~\cite{AR83,PCV88,M98a,
*M98b,S00,MH00,Y02,M04,GM06} and quantum information~\cite{HHH09}.
We find that strong electron correlation, considered as a {\em
unique variable in this model}, dramatically enhances the efficiency
of the energy transfer to the reaction center by more than 100\%.

The results from the multi-electron/exciton models, which unlike
one-electron/exciton models allow us to control the amount of strong
electron correlation in the chromophores, are consistent with the
notion that strong electron correlation is likely employed by nature
to enhance its energy-transfer efficiency as much as other factors
such as (i) environmental noise~\cite{CS09,CCD10,HC10,MN11,
PAM09,ZKR11} and (ii) entanglement between chromophores~\cite{BWV10,
BE11,SIF10}, which have been extensively studied in the recent
literature.  In the multi-electron/exciton model the correlation of
electrons within a molecular subunit like a chromophore is
intrinsically connected with the entanglement of electrons between
molecular subunits.  Furthermore, the results suggest a general
design principle for man-made materials in which electron
correlations and entanglements both within and between subunits are
simultaneously tuned for achieving enhanced quantum efficiencies.
The results give guidelines and inspiration for the engineering of
new materials if their properties can be tailored to match the
parameters of the model.  In combination with other recent advances,
including the study of functional subsystems of the FMO
complex~\cite{SM11}, this interplay of electron entanglements on
different length scales may enable us to develop materials with
quantum efficiencies approaching those found in natural processes
from photosynthesis to bioluminescence.

\section{Theory}

\subsection{Single- and many-electron chromophore models}

The Fenna-Matthews-Olson (FMO) complex of green-sulfur bacteria
contains three identical subunits, each with a network of seven
chromophores.  Very recent crystallographic and quantum chemistry
studies~\cite{SMM11,TWG09} indicate that there is likely an eighth
chromophore in the FMO of {\em Prosthecochloris aestuarii}, which
due to sample preparation is not present in the ultrafast
spectroscopic studies~\cite{ECR07,CWW10,PHF10}.  Theoretical models
of an FMO complex's subunit typically represent each of the seven
chromophore by a {\em one-electron model} in which the electron has
access to two energy levels separated by the excitation energy of
the chromophore. Interactions ${\hat U}$ between pairs of
chromophores are modeled by the exchange of single-electron
excitations (or {\em excitons}) between them:
\begin{equation}
\label{eq:H1} {\hat H} = {\hat H}_{0} + {\hat U}
\end{equation}
where
\begin{eqnarray}
\label{eq:H0} {\hat H}_{0} & = & \frac{1}{2} \sum_{s,m}{ m
\epsilon_{s}
{\hat a}^{\dagger}_{s,m} {\hat a}_{s,m} } \\
\label{eq:U} {\hat U} & = & \sum_{s \neq t}{ U_{s,t} {\hat
a}^{\dagger}_{s,+1} {\hat a}_{s,-1} {\hat a}^{\dagger}_{t,-1} {\hat
a}_{t,+1} }.
\end{eqnarray}
The quantum numbers $s$ and $t$ denote the seven sites of the
chromophores while the quantum number $m$, equal to +1 or -1,
indicates one of the two energy levels within each chromophore.  The
second-quantized operator ${\hat a}^{\dagger}_{s,m}$ (${\hat
a}_{s,m}$) creates (annihilates) an electron on chromophore $s$ in
energy level $m$.  The 7 parameters $\epsilon_{s}$ are the
excitation energies of the chromophore, and the 21 parameters
$U_{s,t}$ are the coupling energies between all pairs of
chromophores.  Typical values for the excitation and coupling
energies are given in the 7x7 Hamiltonian of Ref.~\cite{PH08},
derived from the work of Adolphs and Renger~\cite{renger}.

Some theoretical treatments (see for example Ref.~\cite{SM11})
express this model Hamiltonian in terms of operators ${\hat
b}^{\dagger}_{s}$ and ${\hat b}_{s}$ that create and annihilate an
exciton on chromophore $s$.  By applying the following substitutions
\begin{eqnarray}
{\hat a}^{\dagger}_{s,-1} {\hat a}_{s,-1} & = & {\hat
b}_{s} {\hat b}^{\dagger}_{s} \\
{\hat a}^{\dagger}_{s,+1} {\hat a}_{s,+1} & = & {\hat
b}^{\dagger}_{s} {\hat b}_{s} \\
{\hat a}^{\dagger}_{s,-1} {\hat a}_{s,+1} & = & {\hat b}_{s} \\
{\hat a}^{\dagger}_{s,+1} {\hat a}_{s,-1} & = & {\hat
b}^{\dagger}_{s},
\end{eqnarray}
we can write the electronic Hamiltonian in Eqs.~(\ref{eq:H0}) and
(\ref{eq:U}) as an excitonic Hamiltonian
\begin{eqnarray}
\label{eq:H0e} {\hat H}_{0} & = & \frac{1}{2} \sum_{s}{ \epsilon_{s}
( {\hat b}^{\dagger}_{s} {\hat b}_{s} - {\hat b}_{s} {\hat
b}^{\dagger}_{s} ) } \\ \label{eq:Ue} {\hat U} & = & \sum_{s \neq
t}{ U_{s,t} {\hat b}^{\dagger}_{s} {\hat b}_{t} }.
\end{eqnarray}
Because these mappings are exact, these two Hamiltonians are {\em
equivalent}.  Although the site energies $\epsilon_{s}$ and coupling
energies $U_{s,t}$ will account for electron correlation in an
average fashion if they are determined from quantum calculations
and/or experimental data, these Hamiltonians do not explicitly
correlate the electrons within the chromophores.  The purpose of the
present paper is to design an extension of the one-electron
(exciton) model that treats the electron correlation within the
chromophore explicitly---both in the stationary-state solutions of
the Schr{\"o}dinger equation and in the time-dependent (dynamic)
solutions of the quantum Liouville equation.

Here we generalize this one-electron (exciton) representation by
using the $N$-electron LMG model~\cite{LMG65}.  Each chromophore is
modeled as $N$ electrons in two energy levels that are each $N$-fold
degenerate; a pairwise interaction that excites two electrons from
the lower level to the upper level or de-excites two electrons from
the upper level to the lower level~\cite{M98a}. The Hamiltonian of
the 7 interacting LMG chromophores can be expressed as follows:
\begin{equation}
\label{eq:H2} {\hat H} = {\hat H}_{0} + \lambda {\hat U} + {\hat V}
\end{equation}
where
\begin{eqnarray}
{\hat H}_{0} & = & \frac{1}{2} \sum_{s,m,p}{ m \epsilon_{s}
{\hat a}^{\dagger}_{s,m,p} {\hat a}_{s,m,p} }, \\
{\hat U} & = & \frac{1}{N} \sum_{s \neq t,p}{ U_{s,t} {\hat
a}^{\dagger}_{s,+1,p} {\hat a}_{s,-1,p} {\hat a}^{\dagger}_{t,-1,p}
{\hat a}_{t,+1,p} }, \\ {\hat V} & = & \frac{1}{2} \sum_{s,m,p,q}{ V
| \epsilon_{s} | {\hat a}^{\dagger}_{s,m,p} {\hat
a}^{\dagger}_{s,m,q} {\hat a}_{s,-m,q} {\hat a}_{s,-m,p} },
\end{eqnarray}
where the dimensionless parameter ${\hat V}$ controls the strength
of the electron interactions {\em within} each chromophore and
$\lambda \in [0,1]$ is a dimensionless screening parameter to be
defined below. The product $V | \epsilon_{s} |$ is the interaction
energy for the $s^{\rm th}$ chromophore.  This definition ensures
that for each chromophore the ratio of the interaction energy to the
site energy is the same constant $V$.  The new quantum number $p$
(or $q$) denotes the $N$ degenerate states within each energy level,
which are necessary to accommodate the $N$ electrons.  When the
interaction strength $V$ equals zero, the $N$ electrons on each site
are non-interacting, and with $\lambda=1$ the model behaves the same
as the one-electron-per-site models; when $V$ is unequal to zero, we
have a generalized model for light harvesting with a tunable
electron correlation on the chromophores.

The $N$-electron LMG model has $N+1$ distinct ``molecular orbital''
configurations corresponding to the excitation of zero though $N$
electrons.  If $N$ is even, there are $N/2$ non-degenerate even
excited-state configurations (with an even number of excitations)
and $N/2$ non-degenerate odd excited-state configuration (with an
odd number of excitations). The selection of $N$ in the LMG model,
therefore, depends more on the number of configurations to be mixed
rather than the number of electrons within the chromophore. Choosing
$N=2$ is unsuitable because the single excitations cannot mix with
another odd configuration.  In section~\ref{sec:res} we choose $N=4$
to correlate the single excitations within the chromophore with the
triplet excitations within the chromophore.  While not shown,
similar results are obtained with $N=3$ as well as with $N=5$ and
$N=6$.

\begin{figure}[ht!]

\label{fig:pops}

\begin{center}
\subfigure{\includegraphics[scale=0.3]{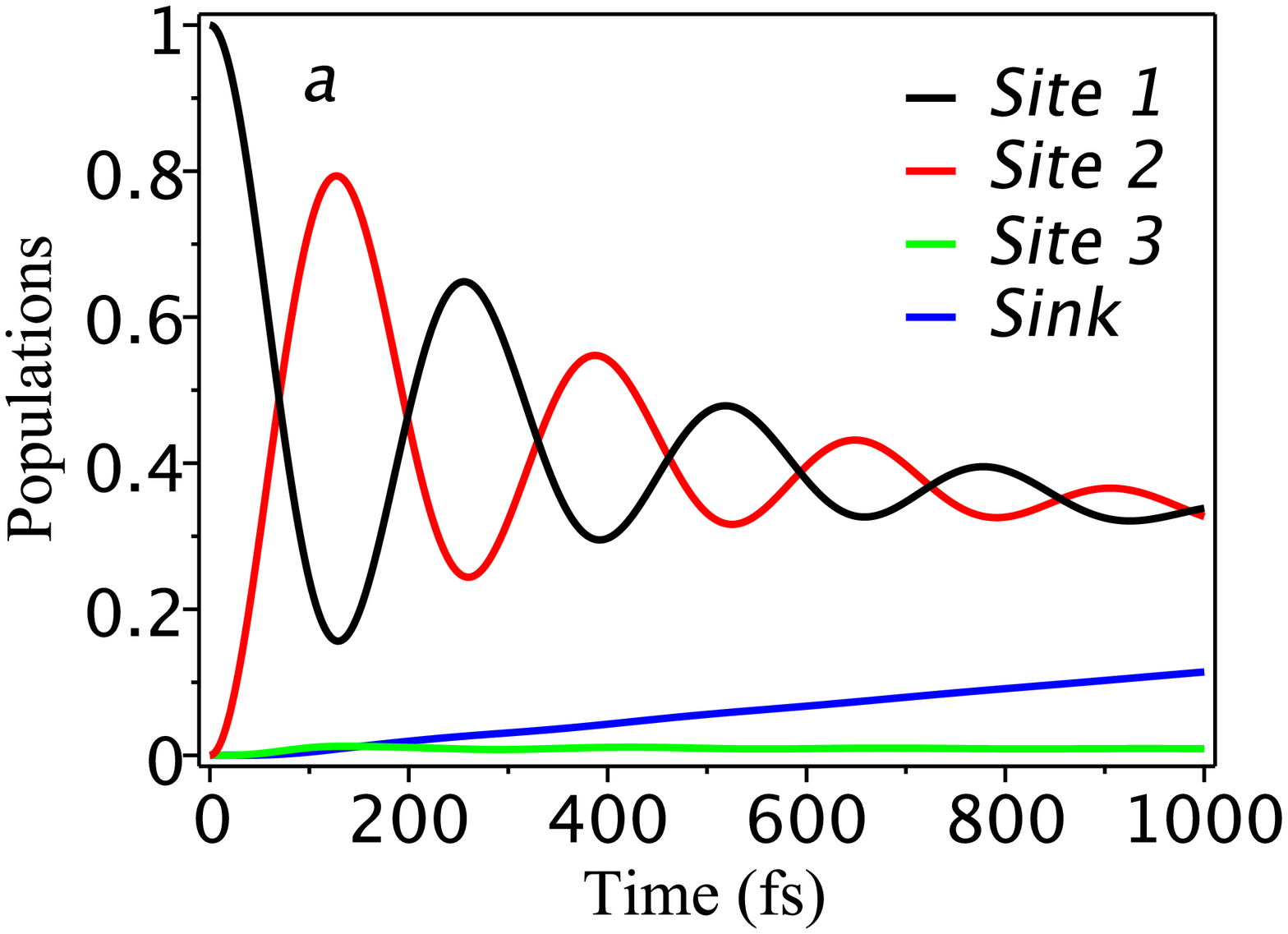}}
\subfigure{\includegraphics[scale=0.3]{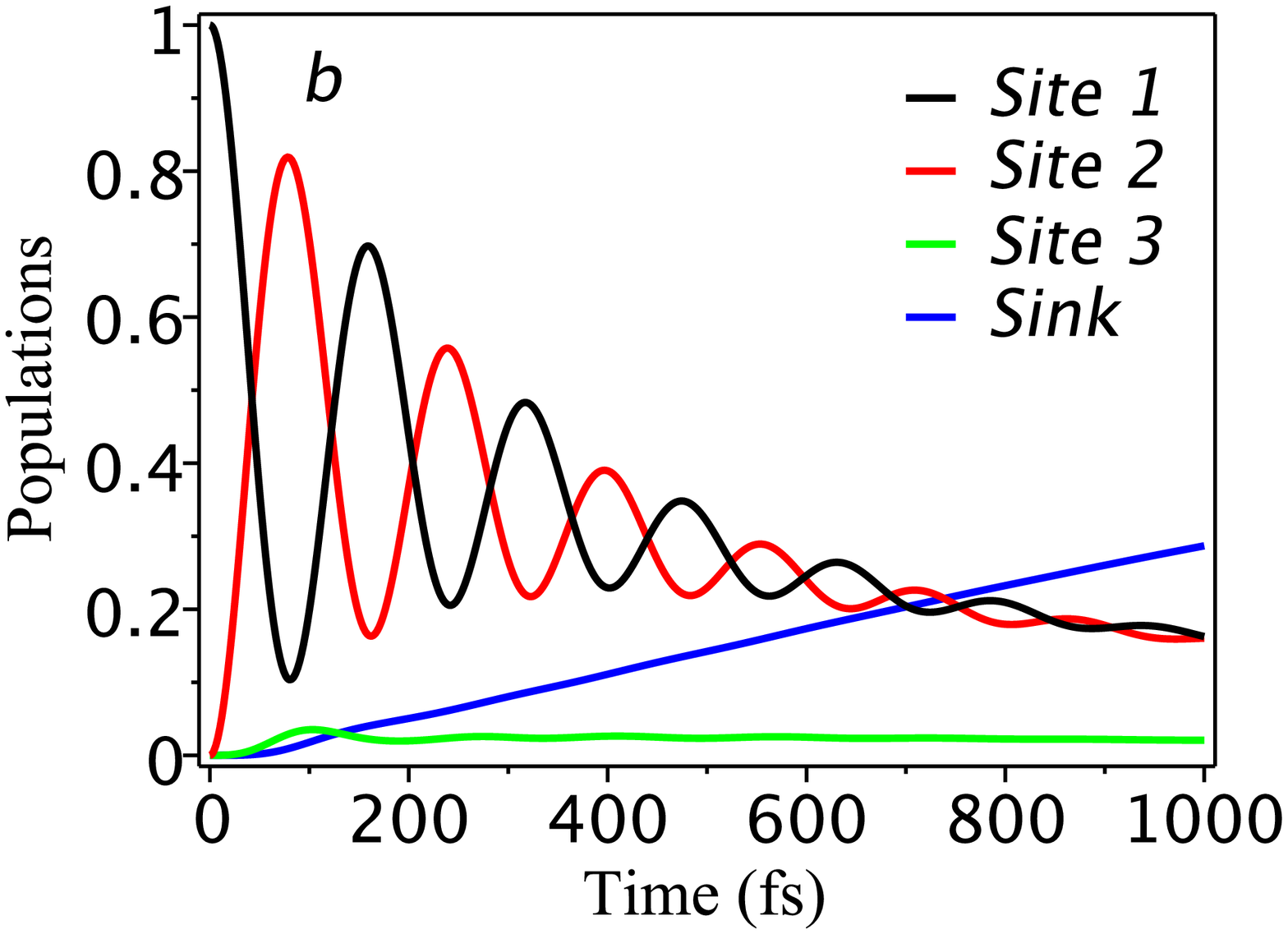}}
\end{center}
\caption{{\bf Populations of chromophores 1-3 and sink with (b) and
without (a) electron correlation per site.} The exciton populations
in chromophores 1, 2, and 3 as well as the sink population are shown
as functions of time in femtoseconds (fs) for $N=4$ and  $\lambda =
0.629$ with (a) $V=0.0$ and (b) $V=0.8$.  Correlating the four
electrons on each chromophore significantly accelerates the increase
in the sink population with time.}

\end{figure}

\subsection{Environmental effects}

Environmental effects of dephasing and dissipation as well as the
transfer of energy to the reaction center (sink) can be incorporated
by introducing a Lindblad operator ${\hat L}$ into the {\em quantum
Liouville equation}
\begin{equation}
\label{eq:L} \frac{d}{dt} D = - \frac{i}{\hbar}[ {\hat H}, D ] +
{\hat L}(D)
\end{equation}
where $D$ is the many-electron density matrix.  The Lindblad
operator can be divided into three operators that account for
dephasing, dissipation, and the sink
\begin{equation}
\label{eq:LB} {\hat L}(D) = {\hat L}_{\rm deph}(D) + {\hat L}_{\rm
diss}(D) +
{\hat L}_{\rm sink}(D) \\
\end{equation}
where
\begin{eqnarray}
{\hat L}_{\rm deph}(D) & = & \alpha \sum_{k}{2 \langle k | D | k
\rangle |k \rangle \langle k |  -\{| k \rangle \langle k |,D\} } ,
\\
{\hat L}_{\rm diss}(D) & = & \beta  \sum_{k}{ 2 \langle k | D | k
\rangle |g \rangle \langle g |  -\{| k \rangle \langle k |,D\} }
, \\
{\hat L}_{\rm sink}(D) & = & 2 \gamma \langle 3 | D | 3 \rangle | s
\rangle \langle s |  - \gamma \{| 3 \rangle \langle 3 |,D\} .
\end{eqnarray}
The state $|g \rangle$ denotes the ground eigenstate of the
Hamiltonian in Eq.~(\ref{eq:H2}), the states $|k \rangle$ represent
the excited eigenstates computed from this Hamiltonian where the
interaction ${\hat U}$ between chromophores is set to zero, the
state $| s \rangle$ denotes the reaction center (sink), and $| 3
\rangle$ indicates the first excited state of the third chromophore
multiplied by the ground states of the other chromophores.
Non-Markovian effects from the environment can also be added to the
quantum Liouville equation (for example, refer to
Ref.~\cite{ZKR11}); however, they will not qualitatively change the
effect of strong electron correlation within chromophores on
energy-transfer efficiency.

\section{Applications}

\subsection{Computations details}


\begin{table}[t!]

\caption{The parameters $\{ \epsilon_{s} \}$, given below, are
chosen to ensure that the excitations of the correlated LMG models
agree with the excitation energies of the seven chromophores from
Ref.~\cite{PH08}.}

\label{t:eps}

\begin{ruledtabular}
\begin{tabular}{cccccccc}

& \multicolumn{7}{c}{Parameters $\epsilon_{s}$ of the LMG Model (cm$^{-1}$)} \\
\cline{2-8} V & $\epsilon_{1}$ & $\epsilon_{2}$ & $\epsilon_{3}$ &
$\epsilon_{4}$ & $\epsilon_{5}$ & $\epsilon_{6}$ & $\epsilon_{7}$ \\
\hline

0.0 & 12450 & 12455 & 12235 & 12360 & 12685 & 12565 & 12515 \\

0.4 & 14293 & 14299 & 14046 & 14190 & 14563 & 14425 & 14368 \\

0.8 & 15227 & 15234 & 14965 & 15117 & 15515 & 15368 & 15307 \\

1.2 & 14201 & 14206 & 13955 & 14098 & 14469 & 14332 & 14275

\end{tabular}
\end{ruledtabular}

\end{table}

The values for the site excitation energies are defined by the 7x7
Hamiltonian of Ref.~\cite{PH08}, which is derived from the data in
Ref.~\cite{renger}; when $V > 0$, the parameters $\{ \epsilon_{s}
\}$ are adjusted, as shown in Table~\ref{t:eps}, to ensure that the
excitations of the correlated LMG models agree with the excitation
energies of the seven chromophores. Based on acene-chain~\cite{GM08,
*PGG11} and metalloporphyrin-ring data~\cite{CFC11}, which imply
that the population of the lowest unoccupied orbital (unoccupied in
a mean-field treatment) is at least 20\%, we estimate the
interaction strength $V$ with $N=4$ to be 0.8. The estimate of $V$
is made by finding the value of $V$ that gives an 80\% probability
of finding an electron in the highest occupied orbital (occupied in
a mean-field treatment) and a 20\% probability of finding an
electron in the lowest unoccupied orbital. Lower values of $V$ give
less correlation and hence, less depletion of the highest mean-field
occupied orbital.  This estimate is conservative because: (i) the
acene chains of a similar length typically reveal a nearly biradical
filling ($\approx$ 50\% in the highest occupied orbital), and (ii)
the presence of the Mg ion with its $d$ orbitals is expected to
enhance the degree of correlation.

Because the coupling energies $U_{s,t}$ given in Ref.~\cite{PH08}
are ``dressed'' dipole-dipole interactions that account implicitly
for both the electron correlation within the chromophores and the
protein environment surrounding the chromophores, they require
adjustment for the LMG chromophore model that contains an explicit
electron-electron interaction.  To prevent overcounting of the
electron correction's effect on the coupling, we screen the coupling
energies $U_{s,t}$ of Ref.~\cite{PH08} by selecting $\lambda$ in
Eq.~(\ref{eq:H2}) to be less than unity.  Specifically, we set
$\lambda = 0.629$ in {\em all} of the calculations reported in
section~\ref{sec:res} to match experimental transfer efficiency to
the sink with the LMG model when $N=4$ and $V=0.8$. {\em Without
screening} the transfer efficiency at $V=0.8$ (and $\lambda = 1$) is
observed to be significantly greater than its experimental value
because the electron correlation is counted {\em twice}: (i)
implicitly in the coupling energies $U_{s,t}$ and (ii) explicitly in
the LMG model of the chromophores. The rate parameters $\alpha$,
$\beta$, and $\gamma$ in the Lindblad operators in Eq.~(\ref{eq:LB})
are chosen in atomic units (and wavenumbers) to be $1.52 {\rm x}
10^{-4}$~a.u. (33.4~cm$^{-1}$), $7.26 {\rm x} 10^{-5}$~a.u.
(15.9~cm$^{-1}$), and $1.21 {\rm x} 10^{-8}$~a.u.
(0.00266~cm$^{-1}$), respectively.  These definitions are similar to
those employed in previous work with the one-electron/exciton
model~\cite{PH08}.

\subsection{Results}

\label{sec:res}

The exciton populations in chromophores 1, 2, and 3 as well as the
sink population are shown as functions of time in femtoseconds (fs)
in Figs.~2a and~2b for $N=4$ and $\lambda = 0.629$ with $V=0.0$ and
$V=0.8$, respectively. Population dynamics of the excitation are
generated by evolving the Liouville equation in Eq.~(\ref{eq:L})
from an initial density matrix with chromophore~1 in its first
excited state and the other chromophores in their ground states.
Correlating the 4 electrons on each chromophore significantly
accelerates the increase in the sink population with time.  By 1~ps
the sink population for $V=0.0$ is 0.114 while the population for
$V=0.8$ is 0.287. Correlating the excitons on each chromophore also
has the effect of shortening the periods of oscillation in
chromophores~1 and~2 and accelerating the population decay in these
chromophores, which is consistent with the change in the sink
population.  The population dynamics with $N=4$, $\lambda = 0.629$,
and $V=0.8$ in Fig.~2b can also be compared with the dynamics of the
one-electron (exciton) model Hamiltonian (equivalent to $N=4$,
$V=0$, and $\lambda=1$), shown in Fig.~1b of Ref.~\cite{SM11}.  The
two figures are nearly identical including the efficiency with which
energy is transferred to the sink.  Hence, the LMG model of the
chromophores matches the experimental results but with an explicitly
correlated treatment of the electrons on the chromophores.

The sink population as a function of time (fs) is shown in Figs.~3
for a range of $V$ with $N=4$ and $\lambda = 0.629$.  Importantly,
as $V$ increases, we observe a dramatic acceleration of the increase
of the sink population.  For $V$ increasing by the sequence 0.0,
0.4, 0.8, and 1.2, the sink population at 2~ps increases by the
sequence 0.221, 0.367, 0.498, and 0.547.  For $N=4$ correlating the
electrons within the chromophores significantly increases the
efficiency of energy transfer to the reaction center (sink) by as
much as 148\%.  While we choose $N=4$, the number $N$ of electrons
per chromophore can model electron correlation for any $N > 2$.  The
precise value of $N > 2$ is unimportant because the effect of
changing $N$ can be related to a rescaling of the interaction $V$.

\begin{figure}[ht!]

\label{fig:sinks}

\begin{center}
\subfigure{\includegraphics[scale=0.3]{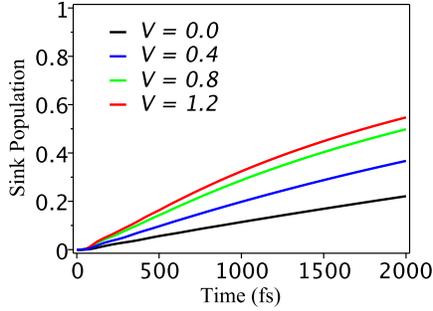}}
\end{center}
\caption{{\bf Correlation-enhanced transfer to the reaction center.}
The reaction center (sink) population as a function of time (fs) is
shown in for a range of $V$ with $N=4$ and $\lambda = 0.629$.
Correlating the electrons within the chromophores significantly
increases the efficiency of energy transfer to the reaction center
(sink).}

\end{figure}

Figure~4 examines the entanglement of excitons between the LMG-model
chromophores for $N=4$ and $\lambda = 0.629$ with $V=0.0$ and
$V=0.8$. We employ a measure of global entanglement in which the
squared Euclidean distance between the quantum density matrix and
its nearest classical density matrix is
computed~\cite{PR02,BNT02,H78,MH00}:
\begin{equation}
\sigma(D)  =  || D - C ||^{2} =  \sum_{i,j}{ |D^{i}_{j} -
C^{i}_{j}|^2 } .
\end{equation}
In some cases like the entanglement of the chromophores, the squared
Euclidean distance can be viewed as the sum of the squares of the
{\em concurrences}~\cite{K07b}, a measure of local entanglement.
Within the mathematical framework of Bergmann distances, the squared
Euclidean distance can also be related to {\em quantum relative
entropy}~\cite{HHH09,SIF10}, which is often applied as a global
entanglement measure.  The squared Euclidean distance $\sigma(D)$ is
nonzero if and only if the excitons on different chromophores are
entangled.  The correlation of electrons increases the degree of the
entanglement between chromophores at early times and the frequency
of its oscillation. The greater entanglement at early times reflects
the opening of additional channels between chromophores for quantum
energy transfer.

\begin{figure}[ht!]

\label{fig:ent}

\begin{center}
\includegraphics[scale=0.3]{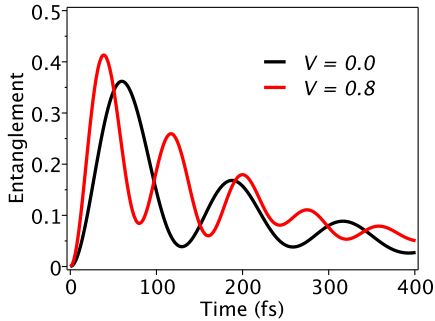}
\end{center}

\caption{{\bf Entanglement of excitons with and without electron
correlation.} A measure of global entanglement is shown as a
function of time (fs) for $N=4$ and $\lambda = 0.629$ with $V=0.0$
and $V=0.8$. The correlation of the excitons increases the degree of
the entanglement between chromophores at early times and the
frequency of its oscillation.}

\end{figure}

\section{Discussion and Conclusions}

\label{sec:con}

The chromophores interact through intermolecular forces, both
dipole-dipole and London dispersion forces.  The correlated-model
results presented here are consistent with the notion that nature
enhances these intermolecular forces through strong electron
correlation in the $\pi$-bonded networks of the chromophores to
achieve the observed energy-transfer efficiency. The two parameters
of the LMG chromophore model provide the simplest approach to
studying the effect of strong electron correlation $V$ on the
effective coupling between chromophores.  The one-electron or dipole
models with their coupling energies $U_{s,t}$ can mimic the
efficiency from such correlation within chromophores through an
empirical inflation of the one-electron $U$ coupling, but they do
not provide a mechanism for either isolating or estimating the
magnitude of the enhanced coupling due to strong electron
correlation.  Orbital occupations from recent 2-RDM calculations of
correlation in polyaromatic hydrocarbons~\cite{GM08,*PGG11} suggest
$V = 0.8$ to be a conservative estimate of the correlation within
the LMG models of the 7 chromophores.  Using this estimate with
coupling energies screened to match experimental and computational
data, we project a greater than 100\% enhancement from the strong
electron correlation relative to the theoretical limit in which
electron correlation within the chromophore is absent.

Correlation-enhanced energy transfer can be compared with
noise-assisted transfer.  Theoretical models~\cite{CS09,CCD10,
HC10,MN11,PAM09,BE11,SIF10} have shown that noise from the
environment (dephasing) can assist energy transfer in the FMO
complex by interfering with the coherence (resonance) between
chromophores with similar energies, which facilitates the downhill
flow of energy to the lowest-energy, third chromophore, connected to
the reaction center.  Electron correlation on each chromophore, we
have shown, enhances transfer by opening additional coherent
channels between chromophores, which also accelerates energy
transfer to the third chromophore.  Photosynthesis can draw from
both of these sources, strong electron correlation within
chromophores and environmental noise, to increase the rates of
energy transfer.  Some of the experimental energy-transfer
efficiency attributed to noise in one-electron chromophores models
may in fact be due to strong electron correlation.

Briggs and Eisfeld~\cite{BE11} recently examined whether the
energy-transfer efficiency from quantum entanglement might be
matched by a purely classical process.  They conclude that if
chromophores are approximated as dipoles, then quantum and classical
treatments can achieve similar efficiency.  While their result might
also be extendable to other dipole or one-electron approximations of
the chromophores, many-electron models of the chromophores cannot be
represented within classical physics.  Neither the electron
correlation, present in the LMG model of the chromophores when
$V>0$, nor the associated enhancement of energy-transfer efficiency
can be mapped onto an analogous classical process.

Many theoretical models~\cite{CS09,CCD10,HC10,MN11,PAM09,BE11,
SIF10} have been designed to explore the energy transfer in recent
light-harvesting experiments, but most of them treat each
chromophore by a single electron with two possible energy states. In
reality, however, the chromophores are assembled from chlorophyll
molecules that contain an extensive network of conjugated
carbon-carbon bonds surrounding magnesium ions, from which
significant strong electron correlation, including polyradical
character, has been shown to emerge~\cite{HDA07,GM08}.  Recent
experimental efforts, not yet published, provide further insights
into the model developed here.  Light-harvesting devices with
artificial acene-like chromophores are being synthesized and shown
to exhibit long-lived coherence.  Once the details of these
experiments are released, the present multi-electron model and
future extensions will be useful in evaluating the role of strong
electron correlation in these devices.  Furthermore, the present
model predicts that the coupling between artificial chromophores,
and hence their energy-transfer efficiency, can be enhanced by
changing their electronic structure to increase their electron
correlation. Experimentally, this prediction can be tested through
functional group substitutions or more fundamental changes in the
synthetic chromophores that systematically modulate the degree of
electron correlation.  Based on the present model, we predict that
significant changes in coupling and efficiency, statistically
related to the amount of electron correlation, will be observed.

In this paper we have examined the effect of strong electron
correlation and entanglement {\em within} chromophores through an
extension of single-electron/exciton models of the chromophores to
$N$-electron models, based on the LMG model~\cite{LMG65,
M98a,MH00,HHH09}. We find that increasing the degree of electron
correlation of each LMG-model chromophore significantly enhances the
efficiency with which energy is transferred to the reaction center
(sink).  This model-based result is consistent with the notion that
nature likely uses strong electron correlation to achieve its
energy-transfer efficiency.  By mixing and yet controlling the
interplay of electron correlation within and between subunits, the
model also has implications for the design of more energy- and
information-efficient materials.

\begin{acknowledgments}

DAM gratefully acknowledges the ARO Grant No. W91 INF-l 1-1-0085,
NSF CAREER Grant No. 0644888, Microsoft Corporation, Dreyfus
Foundation, and David-Lucile Packard Foundation for support.

\end{acknowledgments}


%

\end{document}